\documentclass[11pt]{article}
\usepackage{amsfonts}
\usepackage{latexsym}

\newcommand{\sect}[1]{\setcounter{equation}{0}\section{#1}}

\def\be{\begin{equation}}
\def\ee{\end{equation}}
\def\bea{\begin{eqnarray}}
\def\eea{\end{eqnarray}}

\def\1{\'{\i}}                           
\def\R{{\mathbb R}}

\def\>#1{{\bf #1}}

\def\k{\omega}

\def\mink{{\cal CM}^{3+1}}
\def\gal{{\cal CG}^{3+1}}
\def\euc{{\cal CE}^{4}}

\def\nuevo{l}

\def\diag{\mbox{diag\,}}

\def\opp{{\hat {\mathtt  p}}}
\def\opx{{\hat {\mathtt  x}}}

\begin{document}

\thispagestyle{empty}

\ 
\hfill\

\begin{center}

{\Large{\bf{New
quantum  (anti)de Sitter algebras and}}} 

{\Large{\bf{ discrete symmetries}}}

\end{center}

\bigskip 
\bigskip

\begin{center}   
 Francisco J. Herranz 
\end{center}

\begin{center} {\it { 
Departamento de F\1sica,
Escuela Polit\'ecnica Superior\\ 
Universidad de Burgos,
09006 Burgos, Spain }}\\
e-mail: fjherranz@ubu.es
\end{center}

\bigskip 
\bigskip

\begin{abstract}
\noindent
Two  new quantum anti-de Sitter $so(4,2)$ and
de Sitter $so(5,1)$ algebras are presented. These deformations are
called either `time-type' or `space-type' according to the 
dimensional properties of the deformation parameter. Their  Hopf
structure, universal $R$ matrix and  differential-difference
realization  are obtained in a unified setting by considering a 
contraction parameter  related to the speed of light, which ensures a
well defined non-relativistic limit. Such quantum algebras are shown
to be  symmetry algebras of either  time or space discretizations of 
  wave/Laplace equations on  uniform   lattices. These results lead to
a proposal for   time  and space discrete Maxwell equations with
quantum algebra symmetry.
\end{abstract}

\newpage


\sect{Introduction}

Quantum deformations of the Poincar\'e algebra have been studied
during last years in a search of   more general symmetries that
could   lead to new relativistic theories at the Planck scale.
Amongst them,   we remark the well known   $\kappa$-Poincar\'e
algebra \cite{Lukierskia,Giller,Lukierskib} (of Drinfeld--Jimbo type)
and the so called null-plane quantum Poincar\'e   algebra
\cite{nulla,nullb} (of non-standard or triangular type). The former
is currently under  a great research activity that explores different
traits  as, for instance,     wave functions and free fields on the
associated $\kappa$-Minkowskian space \cite{Kosinski}, 
  boost transformations  \cite{Kowalski,Bruno},  
$\kappa$-deformed electrodynamics \cite{Cougo} and   Hawking
radiation \cite{Blaut}.

Nevertheless, any  quantum Poincar\'e symmetry  should be taken as an
intermediate stage in the construction of   more general structures
such  as   quantum deformations of the conformal or anti-de Sitter
algebra $so(4,2)$. Natural properties expected for such possible
quantum conformal algebras should include, at least,   a well defined
non-relativistic limit to a quantum conformal Galilean algebra, the
existence of some kind of Poincar\'e Hopf subalgebra  as well as a
clear dimensional interpretation of the deformation parameter.  The
aim of this paper  is to present  two new non-standard quantum
deformations of
$so(4,2)$  fulfilling the above requirements.  Furthermore
these structures  are  more manageable than other known
non-standard quantum conformal  algebras~\cite{Lukierskic,vulpi} and,
by construction,  properties  known for lower dimensional cases (such
as $so(2,2)$
\cite{vulpiB})  can  be extended to the present dimension.

In the next section, we give a  unified description of the three
 Lie algebras we shall deal with: $so(4,2)$, $so(5,1)$  and their
limit to the conformal Galilean algebra.  The Hopf structure of the
first type of deformations is introduced in section 3; this is   called
`time-type' as the deformation parameter has dimensions of time. 
Their role as discrete symmetries on a uniform time lattice is
explicitly shown in section 4 through a differential-difference
realization.  In this way, we obtain a time discretization of
conformal invariant equations such as wave or massless Klein--Gordon,
Laplace and Maxwell equations.  A parallel procedure is performed for
`space-type' deformations   in the last section.


\sect{Conformal Lie algebras}

The Lie algebras of the groups of conformal transformations 
of the  $(3+1)$D Minkowskian and Galilean  spacetimes as well as of 
the   4D Euclidean space can be studied simultaneously   by means of
a  single  real contraction parameter $\k$; they are  denoted
collectively  $so_\k(4,2)$. These  are spanned by   generators of
rotations $J_i$,  time   $P_0$ and space $P_i$ translations,  boosts
$K_i$,  special conformal transformations $C_\mu$  and  dilations $D$. 
We  will   assume   sum over repeated indices, latin indices   
$i,j,k=1,2,3$,  greek indices   $\mu,\nu=0,1,2,3$, and  three 
components of a generator    will be denoted
  $\>X=(X_1,X_2,X_3)$. The non-vanishing commutation
relations of $so_\k(4,2)$ are given by
\be
\begin{array}{lll}
[J_i,J_j]=\varepsilon_{ijk}J_k ,&\quad
[J_i,K_j]=\varepsilon_{ijk}K_k ,&\quad
[J_i,P_j]=\varepsilon_{ijk}P_k  ,\\[2pt]
[J_i,C_j]=\varepsilon_{ijk}C_k ,&\quad 
[K_i,K_j]=-\k\varepsilon_{ijk}J_k , &\quad
[K_i,P_i]= \k   P_0  ,\\[2pt]
[K_i,P_0]=P_i ,&\quad  [K_i,C_0]=C_i ,&\quad
[K_i,C_i]= \k   C_0 ,\\[2pt]
[P_0,C_0]=-2 D ,&\quad [P_0,C_i]=2 K_i ,&\quad 
[C_0,P_i]=2 K_i  ,\\[2pt] 
   [P_i,C_j]=2 \k( \delta_{ij}D    
-   \varepsilon_{ijk} J_k)  ,&\quad     
[D,P_\mu]=P_\mu ,&\quad  [D,C_\mu]=-C_\mu .
\end{array}
\label{ba}
\ee
  Each specific   Lie algebra is recovered  from $so_\k(4,2)$ once the
contraction   parameter $\k$ is particularized to a real value as
follows:

\noindent
$\bullet$ $so(4,2)\equiv \mink$ for $\k>0$, is  the
conformal algebra   of the $(3+1)$D Minkowskian spacetime, or  
$(4+1)$D anti-de Sitter algebra. The contraction parameter is related
to the speed of light $c$ through $\k=1/c^2$.

\noindent
$\bullet$ $so(5,1)\equiv \euc$ for $\k<0$, is the
conformal algebra of the 4D Euclidean space, or  $(4+1)$D  de
Sitter algebra; $P_0$ should be  considered as another generator of
space translations and $\>K$  as generators of rotations
($\{\>J,\>K\}$ span an $so(4)$ subalgebra).  
  
\noindent
$\bullet$ $t_9(so(3)\oplus so(2,1))\equiv \gal$ for  $\k=0$, where  
$so(3)=\{\>J \}$, $so(2,1)=\{P_0,C_0,D\}$ and  $t_9=\{\>K,\>P,\>C \}$.
This  case corresponds to the conformal algebra of the $(3+1)$D
Galilean spacetime~\cite{confgal} obtained from $\mink$ through the
non-relativistic limit
$c\to\infty$.

As is well known $so_\k(4,2)$ has two 
remarkable Lie subalgebras:    $\{\>J,\>K,\>P,P_0 \}$ that generate
the (kinematical) algebra of isometries of  the space, and
$\{\>J,\>K,\>P,P_0,D\}$ that span  the  Weyl (or    similitude)
subalgebra ${\cal W}_\k$.

 Let us  consider 
 the spacetime coordinates
$x\equiv(x^0,\>x)\equiv(x^0,x^1,x^2,x^3)$ with metric
$(g_{\mu\nu})=\diag(+1,-\k,-\k,-\k)$.   A    vector
field  representation  of $so_\k(4,2)$ reads
\be
\begin{array}{l}
J_i=\varepsilon_{ijk} x^k \partial_j ,\qquad
K_i=-\k x^i\partial_0-x^0\partial_i ,\qquad
 P_\mu=\partial_\mu  ,\\[2pt] 
D=-x^\mu\partial_\mu -1  ,\qquad  
C_0=\left(\k \>x^2 - (x^0)^2 \right)\partial_0+ 2
x^0\left( x^\mu\partial_\mu +1 \right)  ,\\[2pt]
C_i=\left(\k \>x^2-  (x^0)^2 \right)\partial_i-
2\k x^i\left( x^\mu\partial_\mu  +1 \right) ,
 \end{array} 
 \label{bd}
\ee
where $\partial_\mu\equiv \partial/\partial x^\mu$
and $\>x^2=(x^1)^2+(x^2)^2+(x^3)^2$. Under  this realization,  the  
Casimir  of the  kinematical subalgebra    $E= \>P^2  - \k P_0^2 $
gives rise to  
 the following differential   equation  
\be
 (\partial_1^2+\partial_2^2+\partial_3^2-\k\partial_0^2)\Phi
(x) =0 .
\label{bf}
\ee
Since $E$ commutes with
$\{\>J,\>K,\>P,P_0 \}$ and   the remaining 
generators (\ref{bd}) verify 
\be
 [E,D]=-2E ,\qquad
[E,C_0]= 4  x^0 E ,\qquad [E,C_i]= - 4 \k x^i E   ,
 \label{bg}
\ee
we find that all of them are  symmetry operators of (\ref{bf}),
so that $so_{\k}(4,2)$ is the symmetry algebra
of such an  equation. Hence   we  recover the   $(3+1)$D wave  or
massless Klein--Gordon equation    when $\k>0$ \cite{Barut} and the
usual 4D Laplace--Beltrami equation  when   $\k<0$  ($x^0$ should be
seen as another space coordinate instead of time). The    contraction
$\k=0$ gives rise to a  3D  Laplace  equation in the  Galilean 
spacetime; the absence of the time coordinate $x^0$  is in full
agreement with the known non-relativistic electromagnetic theories
that  only allow  {\em static} electric and magnetic limits from the
Maxwell equations \cite{Levy,LeBellac}.


\sect{Time-type quantum  conformal algebras}

Let us consider the non-standard classical $r$
matrix \cite{Drinfelda,Drinfeldb} of
$sl(2,\R)\simeq so(2,1)$ written in a conformal basis
$\{J_3,J_+,J_-\}\equiv 
\{D,P_0,C_0\}$:
\be
r=-\tau D\wedge P_0 ,
\label{ca}
\ee
where $\tau$ is the deformation parameter. 
We now follow the same procedure
applied to  the   $so(2,2)$ case
 \cite{vulpiB}, that is,  we take  (\ref{ca})  as the $r$ matrix for
the whole $so_\k(4,2)$ algebra. The resulting Hopf structure of the
quantum algebra
$U_\tau(so_\k(4,2))$  is  as follows.

\noindent
$\bullet$ Coproduct:
\be
\begin{array}{l}
\Delta(P_0)=1\otimes P_0 + P_0\otimes 1  ,
\qquad \Delta(P_i)=1\otimes P_i + P_i\otimes {\rm e}^{\tau P_0}  
,\\[2pt]
\Delta(J_i)=
1\otimes J_i + J_i\otimes 1 ,\qquad \Delta(D)=1\otimes D  + D\otimes
{\rm e}^{-\tau P_0} ,\\[2pt]
\Delta(K_i)=1\otimes K_i + K_i\otimes 1-\tau D\otimes  {\rm
e}^{-\tau P_0}P_i  ,\qquad \Delta(C_0)=1\otimes C_0 +  C_0\otimes
{\rm e}^{-\tau P_0}  ,\nonumber\\[2pt] 
\Delta(C_i)=1\otimes C_i + C_i\otimes {\rm e}^{-\tau
P_0}+2\tau D\otimes  {\rm e}^{-\tau P_0}K_i-\tau^2 (D^2+D)  \otimes 
{\rm e}^{-2\tau P_0} P_i .
 \end{array} 
 \label{cc}
\ee

\noindent
$\bullet$ Non-vanishing commutation rules that close the Weyl Hopf
subalgebra
$U_\tau({\cal W}_\k)$:
 \be
\begin{array}{lll}
[J_i,J_j]=\varepsilon_{ijk}J_k ,&\qquad
[J_i,K_j]=\varepsilon_{ijk}K_k ,&\qquad
[J_i,P_j]=\varepsilon_{ijk}P_k , \\[2pt]
[K_i,K_j]=-\k\varepsilon_{ijk}J_k ,
 &\qquad   [K_i,P_0]={\rm e}^{-\tau P_0} P_i ,&\qquad
 [D,P_i]=P_i ,\\[2pt]
   \displaystyle{ [K_i,P_i]= \k
   \frac{{\rm e}^{\tau P_0}-1}{\tau} } ,&\qquad
\displaystyle{[D,P_0]=\frac{1-{\rm e}^{-\tau P_0}}{\tau}} .&
\end{array}
\label{cd}
\ee  

\noindent
$\bullet$  Non-vanishing commutation rules that  involve special
conformal transformations:
\be
\begin{array}{ll}
[J_i,C_j]=\varepsilon_{ijk}C_k  ,
 &\qquad [C_0,C_i]=-\tau(D C_i + C_i D) ,\\[2pt]
[K_i,C_0]=C_i  ,&\qquad  [K_i,C_i]= \k (C_0 -\tau D^2) ,\\[2pt]
 [P_0,C_0]=-2 D ,&\qquad [P_0,C_i]=  {\rm e}^{-\tau P_0}
K_i+K_i\,{\rm e}^{-\tau P_0}  ,\\[2pt] 
 [P_i,C_j]=2 \k (\delta_{ij}D    -    \varepsilon_{ijk} J_k) ,&\qquad
[C_0,P_i]=2 K_i+\tau (D P_i + P_i D)  ,\\[2pt] 
 [D,C_i]=-C_i ,&\qquad   [D,C_0]=-C_0+\tau D^2  .
  \end{array}
\label{ce}
\ee

The three quantum algebras included  within $U_\tau(so_\k(4,2))$ are
called `time-type' ones since the deformation  parameter $\tau$ has,
generically, dimension of a {\em time}, with the clear exception of
$U_\tau(so(5,1)) \equiv U_\tau(\euc)$, for which $\tau$ has dimension
of a length ($\tau$ has the inverse dimension to the  translation
generator $P_0$). In this sense, although of a non-standard nature, 
this kind of deformation is close to the $\kappa$-Poincar\'e algebra.

Some relevant subalgebras of
$so_\k(4,2)$   become into Hopf subalgebras  of $U_\tau(so_\k(4,2))$
after deformation. In particular, besides the Weyl subalgebra
$U_\tau({\cal {W}}_\k)\subset U_\tau(so_\k(4,2))$, we find the
following embedding:
\be
U_\tau(so(2,1))\subset U_\tau(so_\k(2,2))\subset  U_\tau(so_\k(3,2))
\subset U_\tau(so_\k(4,2))
\label{ccee}
\ee
with generators $\{D,P_0,C_0\}\subset 
\{D,P_0,P_1,C_0,C_1,K_1\}\subset \dots$ All of these Hopf subalgebras
share the same classical $r$ matrix (\ref{ca}). Notice that the 
kinematical  subalgebras do not become into Hopf subalgebras, but 
the presence of the dilation is essential.   Other kinds of  Hopf
subalgebra embeddings similar to (\ref{ccee}) have been obtained in
\cite{Kulish}.  We also stress that the initial Hopf structure in the
embedding, $U_\tau(so(2,1))$,   underlies the approach to physics at
the Planck scale introduced in~\cite{Majida, Majidb}.

 According to $\k>,=,<0$, the embedding (\ref{ccee}) splits into
three chains  that clearly show the contractions $\k=0$ (vertical
arrows):
\be
\begin{array}{llll}
 U_\tau(so(2,1))&\subset\  U_\tau({\cal CM}^{1+1})
&\subset\ U_\tau({\cal CM}^{2+1})&\subset\ U_\tau(\mink)\cr
 & \quad \qquad \downarrow&
\quad \qquad \downarrow&\quad \qquad \downarrow\\[2pt]
U_\tau(so(2,1))&\subset\ U_\tau({\cal CG}^{1+1} )
&\subset\ U_\tau( {\cal CG}^{2+1}  )&\subset\ U_\tau(\gal)\cr
 & \quad \qquad \uparrow&
\quad \qquad \uparrow&\quad \qquad \uparrow\\[2pt]
 U_\tau(so(2,1))&\subset\ U_\tau({\cal CE}^{2})
&\subset\ U_\tau({\cal CE}^{3})&\subset\ U_\tau(\euc)\cr
\end{array}
\label{cced}
\ee
 The chain (\ref{ccee})   has an important consequence:   properties
previously known for a  low dimensional case can directly be extended
to higher dimensions. A first application is provided by the
universal quantum $R$  matrix of   $U_\tau(sl(2,\R))\simeq
U_\tau(so(2,1))$:
\be
{\cal R}=\exp\{ \tau P_0\otimes D\}\exp\{-\tau D\otimes P_0\}  .
\label{cg}
\ee
By construction, this     element    also gives the universal  $R$
matrix for all the quantum algebras arising in the sequence
(\ref{ccee}).  This result may  further be used in the construction
of   quantum anti-de Sitter and de Sitter spaces  as well as in the 
computation of    differential calculi in such spaces by means of a
matrix realization of $U_\tau(so_\k(4,2))$; for  a quantum anti-de
Sitter space of Drinfed--Jimbo type see   \cite{Chang}.


\sect{Discrete time symmetries}

In this section we  extend 
 the  time discretization of the $(1+1)$D wave    equation
associated to  $ U_\tau(so(2,2))$~\cite{vulpiB}  to  $(3+1)$D.
Commutation rules  (\ref{cd}) and  (\ref{ce}) naturally include {\em
discrete derivatives} through terms as $({\rm e}^{\pm\tau
P_0}-1)/\tau$. Thus if we  take the usual realization of the
translation generators $P_\mu$ as the derivatives $\partial_\mu$, we
obtain a differential-difference realization of  
$U_\tau(so_\k(4,2))$:
\be
\begin{array}{l}
J_i=\varepsilon_{ijk} x^k \partial_j ,\qquad 
K_i=-\k x^i\Delta_0-x^0 T_0^{-1}\partial_i ,\qquad
D=-x^0 T_0^{-1}\Delta_0 -x^j\partial_j -1 ,\\[3pt]
 P_\mu=\partial_\mu  ,\qquad   C_0=\left(\k  \>x^2+ (x^0)^2 T_0^{-1}
\right)\Delta_0+ 2 x^0\left( x^j\partial_j +1 \right)
+\tau\left( x^j\partial_j +1 \right)^2  ,\\[3pt]
C_i=\left(\k  \>x^2 -  (x^0)^2
T_0^{-2}\right)\partial_i- 2\k x^i\left(  x^0 T_0^{-1}\Delta_0+
x^j\partial_j  +1
\right)  +\tau x^0 T_0^{-2}\partial_i  , 
 \end{array} 
 \label{da}
\ee
where we have introduced the  time shift operator $T_0={\rm e}^{\tau
\partial_0}$ and  the discrete derivative in the time direction
$\Delta_0=(T_0-1)/\tau$; these  operators act  on a function
$\Phi(x)\equiv \Phi(x^0,\>x) $ as
\be
T_0  \Phi(x^0,\>x) =\Phi(x^0+\tau,\>x) ,
\qquad
\Delta_0\Phi(x^0,\>x)=\frac{\Phi(x^0+\tau,\>x)-\Phi(x^0,\>x)}{\tau} .
\label{dda}
\ee
  Thefore the deformation parameter $\tau$ is the time lattice
constant on the uniform lattice discretized along $x^0$,  while the
space  coordinates $\>x$ remain as continuous variables.

The Casimir of the  isometries sector   $\{\>J,\>K,\>P,P_0 \}$ turns
out to be
\be
E_\tau= P_1^2+P_2^2+P_3^2   - \k \left(\frac{{\rm e}^{\tau
P_0}-1}{\tau}  
\right)^2  .
\label{db}
\ee
By introducing (\ref{da}) we find a differential-difference equation
given by
\be
 (\partial_1^2+\partial_2^2+\partial_3^2-\k\Delta_0^2)\Phi
(x) =0 .
\label{dc}
\ee
As the   operators (\ref{da}) out of the isometries sector verify
\be
\begin{array}{l}
 [E_\tau,D]=-2E_\tau ,\qquad
[E_\tau,C_0]= 4(   x^0+\tau x^j\partial_j +2 \tau) E_\tau  ,\qquad
[E_\tau,C_i]= - 4 \k x^i E_\tau  ,
 \end{array} 
 \label{dd}
\ee
  we conclude that $U_\tau(so_\k(4,2))$ is the symmetry
algebra of the equation (\ref{dc}).    Each specific quantum algebra,
Weyl subalgebra and associated equation that arise for a particular
value of $\k$ are displayed in table \ref{table1}; the arrows
indicate the contraction $\k=0$ (or
$c\to \infty$) for each (sub)algebra/equation.  The Lie algebra and
continuous picture is recovered when $\tau\to 0$. Note that  $\k=0$
leads to a differential equation with continuous space variables
associated to
$U_\tau(\gal)$, nevertheless   the realization (\ref{da})  is still a
differential-difference one with an intrinsic time discretization.

\begin{table}[ht]
{\footnotesize
 \noindent
\caption{{Time-type quantum conformal  algebras, $U_\tau(so_\k(4,2))$,  
 quantum Weyl  subalgebras $U_\tau( {\cal W}_\k)$ and 
differential-difference equations  according to  
$\k=\{+1,0,-1\}$.}}
\label{table1}
\smallskip
\noindent\hfill
\begin{tabular}{clll}
\hline
\\[-8pt]
$\k$& Quantum conformal algebra  &\ 
Quantum Weyl subalgebra&\quad Differential-difference  equation  \\ 
&   $\{\>J,\>K,\>P,P_0,D,\>C,C_0\}$ &\ $\{\>J,\>K,\>P,P_0,D\}$&
\quad on a uniform time lattice \\[4pt]
\hline
\\[-8pt]
$+1$& Quantum conf.\ Minkowskian &\quad
Quantum Weyl Poincar\'e&\quad Discrete $(3+1)$ wave equation\\
&$U_\tau(\mink)\equiv U_\tau(so(4,2))$  &\quad $U_\tau({\cal WP})$&
\quad $(\partial_1^2+\partial_2^2+\partial_3^2-\Delta_0^2)\Phi=0$
\\[6pt]
&\quad\quad$\downarrow$ &\qquad\
$\downarrow$&\qquad\quad\quad$\downarrow$\\[5pt]
$0$&Quantum conf.\ Galilean  &\quad
Quantum Weyl Galilean&\quad Continuous 3D Laplace equation\\
&$U_\tau(\gal)$ &\quad $U_\tau({\cal WG})$
&\quad $(\partial_1^2+\partial_2^2+\partial_3^2)\Phi=0$
\\[6pt]
&\quad\quad$\uparrow$ &\qquad\
$\uparrow$&\qquad\quad\quad$\uparrow$\\[5pt]
$-1$&Quantum conf.\ Euclidean  &\quad
Quantum Weyl Euclidean&\quad Discrete 4D Laplace equation\\
&$U_\tau(\euc)\equiv U_\tau(so(5,1))$  &\quad $U_\tau({\cal WE})$
 &\quad
$(\partial_1^2+\partial_2^2+\partial_3^2+\Delta_0^2)\Phi=0$\\[4pt]
\hline
\end{tabular}\hfill}
\end{table}

\bigskip

The Hopf structure of
$U_\tau(so_\k(4,2))$ can be
transformed into another one with  classical commutation rules. 
Explicitly, if we consider the non-linear   map defined
by~\cite{vulpiB,Abde}:
\be
{\cal P}_0=\frac{{\rm e}^{\tau P_0}-1}{\tau} ,\qquad
{\cal C}_0=C_0 - \tau D^2 ,
\label{ea}
\ee
with the remaining generators unchanged, 
 we find that  the commutators (\ref{cd}) and (\ref{ce})   are just
the non-deformed ones (\ref{ba}), while the coproduct (\ref{cc}) is
transformed into:
\be
\begin{array}{l}
\displaystyle{ \Delta({\cal P}_\mu)=1 \otimes  {\cal P}_\mu +  {\cal
P}_\mu\otimes  1 + \tau {\cal P}_\mu \otimes  {\cal P}_0
,\qquad  \Delta({\cal J}_i)=1 \otimes  {\cal J}_i +{\cal J}_i\otimes 
1 ,} \\[5pt]
\displaystyle{ \Delta({\cal K}_i)=1 \otimes {\cal K}_i + {\cal
K}_i  \otimes  1 -  {\cal D} \otimes \frac{ \tau {\cal P}_i}{1+
\tau{\cal P}_0}  ,\qquad 
\Delta({\cal D})=1 \otimes  {\cal D} + {\cal
D} \otimes   \frac{1}{1+\tau{\cal P}_0} ,}\\[8pt]
\displaystyle{\Delta({\cal C}_0)=1 \otimes {\cal C}_0 + {\cal
C}_0  \otimes  
\frac{1}{1+\tau{\cal P}_0}-{\cal D} \otimes 
\frac{ 2 \tau }{1+\tau{\cal P}_0}\,{\cal D}  +  ({\cal D}^2+{\cal
D}) \otimes 
\frac{\tau^2  {\cal P}_0}{(1+\tau{\cal P}_0)^2} ,} \\[8pt]
\displaystyle{\Delta({\cal C}_i)=1 \otimes  {\cal C}_i + {\cal
C}_i \otimes   \frac{1}{1+\tau{\cal P}_0}+ {\cal D}  \otimes 
\frac{2 \tau }{1+\tau{\cal P}_0}\,{\cal K}_i -({\cal D}^2+{\cal
D}) \otimes 
\frac{\tau^2   {\cal P}_i}{(1+\tau{\cal P}_0)^2} } .
\end{array}
\label{eb}
\ee
  At the level of
the differential-difference realization (\ref{da}), the
non-linear map    gives rise to
\be
{\cal P}_0=\Delta_0 ,\qquad 
{\cal C}_0=
\left(\k  \>x^2 + (x^0)^2
T_0^{-2}\right)\Delta_0+ 2 x^0 T_0^{-1}\left( x^j\partial_j +1 \right)
-\tau x^0 T_0^{-2}\Delta_0 ,
\label{ec}
\ee
with the remaining
operators and equation (\ref{dc}) unchanged (the latter  now comes
from the non-deformed Casimir $E={\cal P}^2-\k{\cal P}_0^2$).  This
final form for the realization of $U_\tau(so_\k(4,2))$  allows us
to define  `time-type' momenta $\opp_\mu(\partial,x)$ and position
operators
$\opx^\mu(\partial,x)$ as
\be
\opp_0=\Delta_0,\qquad \opp_i=\partial_i,\qquad
\opx^0=x^0 T_0^{-1},\qquad \opx^i=x^i, 
\label{ed}
\ee
that fulfil
\be
[\opp_\mu, \opx^\nu]=\delta_{\mu\nu},\qquad 
[\opp_\mu, \opx_\nu]=g_{\mu\nu},\qquad
[\opx^\mu, \opx^\nu]=[\opp^\mu, \opp^\nu]=0,
\label{ee}
\ee
provided that $[\Delta_0,  x^0 ]=T_0$. 
If we now apply the maps 
$\partial_\mu \longmapsto \opp_\mu$, $x^\mu\longmapsto \opx^\mu$
to the differential
realization (\ref{bd}) and equation (\ref{bf})  
of $so_\k(4,2)$, we recover the  differential-difference realization
(\ref{ec}) (with (\ref{da}) for the remaining operators) and 
equation (\ref{dc}) of 
$U_\tau(so_\k(4,2))$. In this way, we obtain a  kind of {\em
 time-type discretization rule} for quantum algebras. This is rather
similar  to the procedure used  by Aizawa \cite{aizawa01a,aizawa01b}
to deduce a massless Klein--Gordon equation from the non-standard
quantum $so(3,2)$ algebra  given in~\cite{vulpi}.

Furthermore, as  $SO(4,2)$ is the maximal
invariance group of the vacuum Maxwell equations,  
some `discretized' version should be related to $U_\tau(so(4,2))$. 
As the latter is  a   Hopf symmetry algebra  of 
a differential-difference wave equation  with  underlying operators
(\ref{ed}),  it seems that a natural ansatz for  the time 
discretization  of the vacuum Maxwell equations is given by:
\be
\begin{array}{l}
\nabla\cdot\>E=0,\qquad \nabla\cdot\>B=0,\qquad 
\nabla\times \>E=-\Delta_0(\>B),\qquad 
\nabla\times \>B=\frac 1{c^2} \Delta_0(\>E) , 
\label{maxtime}
\end{array} 
\ee
where $\>E$,  $\>B$ are  electric and magnetic fields,
 and
$\nabla=(\opp_1,\opp_2,\opp_3)=(\partial_1,\partial_2,\partial_3)$.
Consistency with the discrete wave equation (\ref{dc}) (with
$\k=1/c^2$)  is guaranteed, since equations (\ref{maxtime}) lead to a
$\tau$-d'Alembertian
$\Box_\tau=\nabla^2 -\frac 1{c^2}
\Delta_0^2$, such that $\Box_\tau \>E=0$ and $\Box_\tau \>B=0$.


\sect{Space-type   quantum conformal  algebras and discrete\\ space
symmetries}

The non-standard classical $r$
matrix of $sl(2,\R)$ can alternatively be written  in the   conformal
basis $\{J_3,J_+,J_-\}\equiv   \{D,P_1,C_1\}$ as $r=-\sigma D\wedge
P_1$,  where $\sigma$ is now the deformation parameter. Hence we
interchange the role of the generators $P_0\leftrightarrow P_1$, so
that $\sigma$  has dimensions of a {\em length}. The resulting Hopf
structure of the `space-type'  quantum algebras  
$U_\sigma(so_\k(4,2))$ is characterized by:

\noindent
$\bullet$ Coproduct:
\be
\begin{array}{l}
 \Delta(P_\mu)=1\otimes P_\mu +P_\mu\otimes {\rm e}^{\sigma P_1} -
\delta_{1\mu} P_\mu\otimes ({\rm e}^{\sigma P_1}-1) ,\\[2pt]
 \Delta(K_j)=1\otimes K_j + K_j\otimes 1-\k \sigma \,\delta_{1j}
D\otimes  {\rm e}^{-\sigma P_1}P_0 ,\\[2pt]
 \Delta(J_j)=1\otimes J_j + J_j\otimes 1+\sigma\, \varepsilon_{1jk} 
D\otimes  {\rm e}^{-\sigma P_1}P_k  ,\qquad  \Delta(D)=1\otimes D +
D\otimes {\rm e}^{-\sigma P_1} ,\\[2pt]
\Delta(C_0)=1\otimes C_0 + C_0\otimes {\rm e}^{-\sigma
 P_1}-2\sigma D\otimes  {\rm e}^{-\sigma P_1}K_1+\k\sigma^2 (D^2+D)
\otimes  {\rm e}^{-2\sigma P_1} P_0 ,\\[2pt]
\Delta(C_j)=1\otimes C_j + C_j\otimes {\rm e}^{-\sigma
P_1}-2\k\sigma \, \varepsilon_{1jk}   D\otimes  {\rm
e}^{-\sigma P_1} J_k\\[2pt]
 \qquad\qquad\qquad +\k\sigma^2(\delta_{2j}+\delta_{3j}) (D^2+D)
\otimes  {\rm e}^{-2\sigma P_1} P_j  .  
 \end{array} 
 \label{fc}
\ee

\noindent
 $\bullet$ Non-vanishing commutation rules closing a Weyl Hopf
subalgebra
$U_\sigma({\cal W}_\k)$:
 \be
\begin{array}{l}
[J_i,J_j]=\varepsilon_{ijk}J_k  ,\qquad
[J_i,K_j]=\varepsilon_{ijk}K_k  ,\qquad
[K_i,K_j]=-\k\varepsilon_{ijk}J_k    ,\\[4pt]
{ \displaystyle{ [J_i,P_j]=\varepsilon_{ijk}\left( \delta_{1i} P_k +
 \delta_{1j}  {\rm e}^{-\sigma P_1}   P_k+
\delta_{1k} \frac{ {\rm e}^{\sigma P_1}-1} {\sigma}    \right)}}
,\\[10pt]
{ \displaystyle{ [K_i,P_i]=\k  P_0\left( 1+  \delta_{1i}({\rm
e}^{-\sigma P_1}-1) \right) }} ,\\[4pt]
{ \displaystyle{ [K_i,P_0]=P_i+\delta_{1i}\left(\frac{ {\rm e}^{\sigma
P_1}-1}{\sigma}-P_1 \right) }   },\\[4pt]
 { \displaystyle{ [D,P_\mu]=P_\mu+\delta_{1\mu}\left(\frac{ 1-{\rm
e}^{-\sigma P_1}}{\sigma}-P_1 \right) } }.
\end{array}
\label{fd}
\ee

\noindent
$\bullet$ Non-vanishing commutation rules involving special conformal
transformations:
\be
\begin{array}{l}
[J_i,C_j]=\varepsilon_{ijk}(C_k+ \k \sigma \,\delta_{1k} D^2) , \qquad
 [C_1,C_\nu]= \k \sigma  (D C_\nu + C_\nu D),\quad \nu\ne 1,\\[2pt] 
[K_i,C_0]=C_i+ \k\sigma\,\delta_{1i} D^2 ,\qquad  [K_i,C_i]= \k  
C_0   , \\[2pt] 
 [P_i,C_j]= -\k \varepsilon_{ijk}\left(
\delta_{1i}({\rm e}^{-\sigma P_1} J_k+J_k\, {\rm e}^{-\sigma
 P_1})+\delta_{1j}(2J_k-\sigma \varepsilon_{ijk}(DP_i+P_i D)) \right)
\\[2pt]
 \qquad\qquad\qquad -2 \k \varepsilon_{ij1}J_1 +2\k \delta_{ij} D
,\\[2pt]
   [C_0,P_i]= 2 K_i+\delta_{1i} \left({\rm e}^{-\sigma P_1}
K_1+K_1\,{\rm e}^{-\sigma P_1}-2 K_1 \right)  ,\qquad [P_0,C_0]=-2 D
,\\[2pt] [P_0,C_i]=2 K_i+\k
\sigma\,\delta_{1i} (D P_0 + P_0 D) ,\qquad  
[D,C_\mu]=-C_\mu-\k\sigma\,\delta_{1\mu} D^2  .
  \end{array}
\label{fe}
\ee

Chains of Hopf subalgebras similar  to the time-type ones
 (\ref{ccee}) and (\ref{cced}) arise  within
$U_\sigma(so_\k(4,2))$.    All of them share the same  universal
$R$  matrix  given by
\be
{\cal R}=\exp\{ \sigma P_1\otimes D\}\exp\{-\sigma D\otimes P_1\}  .
\label{fg}
\ee

In this case, the non-linear map defined by
\be
{\cal P}_1=\frac{{\rm e}^{\sigma P_1}-1}{\sigma} ,\qquad
{\cal C}_1=C_1 +\k \sigma D^2 ,
\label{gf}
\ee
 keeping the remaining generators unchanged,  allows us to rewrite
the Hopf structure of $U_\sigma(so_\k(4,2))$ with non-deformed
commutation rules (\ref{ba}) and  coproduct given by 
\be
\begin{array}{l}
 \displaystyle{\Delta({\cal J}_j)=1 \otimes  {\cal J}_j +{\cal
J}_j\otimes  1+\sigma \varepsilon_{1jk}{\cal D}\otimes \frac{{\cal
P}_k}{1+\sigma{\cal P}_1}} ,\\[10pt]
\displaystyle{ \Delta({\cal K}_j)=1 \otimes {\cal K}_j + {\cal
K}_j   \otimes  1 -  \k \sigma \delta_{1j}{\cal D} \otimes
\frac{{\cal P}_0}{1+
\sigma{\cal P}_1} },\\[10pt]
\displaystyle{\Delta({\cal P}_\mu)=1 \otimes  {\cal P}_\mu +  {\cal
P}_\mu\otimes  1 + \sigma {\cal P}_\mu \otimes  {\cal P}_1,\qquad
\Delta({\cal D})=1 \otimes  {\cal D} + {\cal
D} \otimes   \frac{1}{1+\sigma{\cal P}_1}} ,\\[10pt]
\displaystyle{\Delta({\cal C}_0)=1 \otimes {\cal C}_0 + {\cal
C}_0  \otimes  
\frac{1}{1+\sigma{\cal P}_1}-  {\cal D} \otimes 
\frac{2 \sigma}{1+\sigma{\cal P}_1}\,{\cal K}_1  +  ({\cal D}^2+{\cal
D}) \otimes 
\frac{ \k \sigma^2  {\cal P}_0}{(1+\sigma{\cal P}_1)^2} } ,\\[10pt]
\displaystyle{\Delta({\cal C}_1)=1 \otimes  {\cal C}_1 + {\cal
C}_1 \otimes   \frac{1}{1+\sigma{\cal P}_1}+{\cal D}  \otimes 
 \frac{ 2\k \sigma }{1+\sigma{\cal P}_1}\,{\cal D} -  ({\cal
D}^2+{\cal D}) \otimes 
\frac{\k\sigma^2{\cal P}_1}{(1+\sigma{\cal P}_1)^2} } ,\\[10pt]
 \displaystyle{\Delta({\cal C}_\nuevo)=1 \otimes  {\cal C}_\nuevo +
{\cal C}_\nuevo \otimes   \frac{1}{1+\sigma{\cal P}_1}- {\cal D} 
\otimes 
\frac{2 \k \sigma
\varepsilon_{1\nuevo k}}{1+\sigma{\cal P}_1}\,{\cal J}_k  +  ({\cal
D}^2+{\cal D}) \otimes 
\frac{\k\sigma^2 {\cal P}_\nuevo}{(1+\sigma{\cal P}_1)^2} } ,
\end{array}
\label{gg}
\ee
where   the index $\nuevo=2,3$. Once  $U_\tau(so_\k(4,2))$ is
expressed in this last form,  we can apply a {\em space-type
discretization rule}. Let us consider the space   shift operator
$T_1={\rm e}^{\sigma \partial_1}$ and  the discrete derivative in the
$x^1$-space direction 
$\Delta_1=(T_1-1)/\sigma$. Similarly to the previous section, 
 we define  space-type  momenta  and position
operators, fulfilling (\ref{ee}),  as
\be
\opp_0=\partial_0,\qquad \opp_1=\Delta_1 ,\qquad
\opp_\nuevo=\partial_\nuevo,\qquad
\opx^0=x^0  ,\qquad \opx^1=x^1 T_1^{-1},\qquad \opx^\nuevo=x^\nuevo.
\label{gh}
\ee
By   subsistuting $\partial_\mu \longmapsto \opp_\mu$,
$x^\mu\longmapsto \opx^\mu$ into the differential
realization  (\ref{bd})  of $so_\k(4,2)$, we obtain a
differential-difference realization of $U_\tau(so_\k(4,2))$. The 
associated   differential-difference equation  turns out to be
\be
 (\Delta_1^2+\partial_2^2+\partial_3^2-\k\partial_0^2)\Phi
(x ) =0 .
\label{gk}
\ee
Therefore the deformation  parameter $\sigma$ is the space lattice
constant on the uniform lattice discretized along $x^1$. These
results can be displayed for each particular $\k$ as in table
\ref{table1}.

Finally, if we consider a differential-difference operator
$\nabla_\sigma=(\opp_1,\opp_2,\opp_3)=(\Delta_1,\partial_2,\partial_3)$,
we could make the following ansatz for   a space discretization of
the Maxwell equations  with $U_\tau(so(4,2))$-symmetry:
\be
\begin{array}{l}
\nabla_\sigma\cdot\>E=0,\qquad \nabla_\sigma\cdot\>B=0,\qquad 
\nabla_\sigma\times \>E=-\partial_0 \>B,\qquad 
\nabla_\sigma\times \>B=\frac 1{c^2} \partial_0 \>E , 
\end{array} 
\ee
 which give rise to a $\sigma$-d'Alembertian
$\Box_\sigma=\nabla_\sigma^2 -\frac 1{c^2}
\partial_0^2$, such that $\Box_\sigma \>E=\Box_\sigma \>B=0$.


\section*{Acknowledgments}

This work was partially supported  by the Ministerio de Ciencia y
Tecnolog\1a, Spain (Project BFM2000-1055).  The author thanks Naru
Aizawa for helpful discussions and Susana Garc\1a-Castrillo for
hospitality.


\end{document}